\documentclass{article}[12pt]
\usepackage{doi}
\usepackage{float}

\usepackage[english]{babel}

\usepackage[letterpaper,top=2cm,bottom=2cm,left=3cm,right=3cm,marginparwidth=1.75cm]{geometry}

\usepackage{setspace}

\usepackage{amsmath,amssymb,amsthm,mathrsfs,dsfont,mathtools,euscript,stmaryrd}
\usepackage{graphicx}
\usepackage{authblk}
\graphicspath{{./fig/}{./Old/}}
\DeclareGraphicsExtensions{.pdf,.png,.jpg}

\hypersetup{colorlinks=true, allcolors=blue} 

\usepackage{subcaption}

\newtheorem{theorem}{Theorem}[section]

\newtheorem{lemma}{Lemma}[section]

\usepackage{todonotes}

\title{Dynamics and control of maize infection by \textit{Busseola fusca}: multi-seasonal  modeling and biocontrol strategies}
\author[1]{Clotilde Djuikem\thanks{Corresponding author: \texttt{clotilde.djuikem@umanitoba.ca} (Clotilde Djuikem)}}
\author[2]{Josué Tchouanti}
\date{}

\affil[1]{University of Manitoba, Canada.}
\affil[2]{INRIA Saclay -- Team LIFEWARE, France.}

\begin{document}
\maketitle

\begin{abstract}
Maize production in sub-Saharan Africa faces significant challenges due to the maize stalk borer (\textit{Busseola\,fusca}), a major pest that causes substantial yield losses. Chemical control methods have raised concerns about environmental impact and pest resistance, making biological control a promising alternative. In this study, we develop a multi-seasonal mathematical model using an impulsive system of differential equations to describe stalk borer population dynamics and evaluate pest control strategies. We analyze the stability of the pest-free solution using Floquet theory and study the effects of periodic predator releases on pest suppression. Numerical simulations illustrate the impact of cultural practice and predator release frequency. Moreover, our simulations show that, under good cultural practices, releasing predators once or three times a year is an effective biocontrol strategy. However, in cases of poor cultural practices, biocontrol has only a limited effect, and the best outcome is achieved when predators are released once a year at the beginning of the cropping season.
\end{abstract}



\section{Introduction}

The \textit{Zea mays}, commonly called maize, is one of the most important cereals in the world. This crop satisfied the regular consumption of millions of people and plays an important role in food security \cite{faostat}. However, its production is affected by many pests, particularly the maize stalk borer, \textit{Busseola \,fusca}, which causes significant yield losses across sub-Saharan Africa \cite{ntahomvukiye2018study,fAO}. Common pest control strategies use chemical insecticides which increase resistance and environmental concerns that raises urgent questions about their sustainability \cite{betz2000advantages}.
In response, biological control has been proposed as an eco-friendly alternative, using natural enemies such as parasitoids, predators, and entomopathogenic pathogens to regulate \textit{B.\,fusca} populations \cite{kalule2006effects, harris1992busseola}.
Parasitoids such as \textit{Cotesia sesamiae} have demonstrated effectiveness in controlling \textit{B. fusca} larvae. Studies have reported parasitism rates reaching up to 75\% during peak periods \cite{Calatayud2014}. Similarly, egg parasitoids like \textit{Telenomus busseolae} have been identified as potential biocontrol agents \cite{Calatayud2020}. Predators also play a role in natural pest suppression. Ant species such as \textit{Dorylus helvolus }and rodents like \textit{Mastomys natalensis} have been observed preying on \textit{B. fusca} larvae and pupae, contributing to the reduction of pest population \cite{harris1992busseola}.
Pathogens, including entomopathogenic fungi like \textit{Beauveria bassiana} and bacteria such as \textit{Bacillus thuringiensis,} have shown potential in infecting and killing \textit{B. fusca} larvae, offering a biological alternative to chemical control methods \cite{Calatayud2020,Calatayud2014}.
While biological control has shown promising results, its implementation remains highly variable due to fluctuating ecological conditions and limited predictive tools for optimizing control strategies.
To effectively manage \textit{B. fusca}, we need to understand its population dynamics and the interaction with maize plant.

Mathematical models such as compartmental models \cite{van_den_bosch2007, Djuikem2024_ontogenic} and impulsive differential equations \cite{mailleret2009semi, madden2007study,bainov2017impulsive,LI2015173} have been used to study the dynamics and control of pest populations. These models provide the framework for understanding pest outbreaks, seasonal population variations, and the long-term impact of biological control measures.
In the particular case of \textit{B. fusca }, several studies have developed structured population models to describe its lifecycle stages and their interactions with the environment. Ntahomvukiye et al. \cite{ntahomvukiye2018study} proposed a stage-structured model incorporating larval development, pupation, and adult reproduction, allowing for an estimation of pest population growth under different ecological conditions. Their findings highlighted temperature-dependent variations in \textit{B. fusca} emergence and dispersal patterns, which are critical for optimizing control measures.
Similarly, Tchienkou et al. \cite{TCHIENKOUTCHIENGANG2023379} extended classical pest models by integrating seasonal variations and crop residue management. Their model demonstrated that carryover effects from previous cropping seasons significantly influence infestation levels, emphasizing the need for integrated pest management strategies.
Several studies have explored impulsive control strategies in pest management \cite{Paez2017, TANG2008115,Djuikem_impulsive}. These models capture the effects of periodic pesticide applications, crop harvests, or biological control releases. Djuikem et al. \cite{Djuikem_impulsive} investigated the role of periodic predator releases in pest population suppression.
Despite these advances, there remains a need for models that integrate multi-seasonal pest dynamics, incorporating both impulsive interventions and environmental factors influencing pest survival. This study aims to address these gaps by developing a comprehensive framework for \textit{B.\,fusca} biocontrol using impulsive model.

In this study, we propose a multi-seasonal mathematical model to investigate the dynamics of \textit{B.\,fusca} infestation in a maize field. It is an impulsive system of differential equation that captures past dynamics while taking into account different phases of maize growing. We establish mathematical conditions for pest eradication by analyzing the local and global stability of the pest free periodic solution of this system using the Floquet theory and Lyapunov stability \cite{haddad2006impulsive, bainov2017impulsive}. Numerical simulations are performed to illustrate this result. We also discuss the influence of considering pest immigration at a constant rate, which can represent an alternative host during the non-cropping season.

Further, following the ideas of Djuikem et al.\,\cite{Djuikem_impulsive}, we introduce a biocontrol strategy that consists of releasing a predator of \textit{B.\,fusca} in the field. We investigate the new mathematical condition that ensures pest eradication, in which we highlight the contribution of biocontrol. As expected, it makes it much easier to stabilize the pest-free state, helping to control pest growth. We run numerical simulations to study how control factors like predator release frequency and environmental capacity affect the system.

The paper is structured as follows. In Section~\ref{sec:model-math}, we present the multi-seasonal mathematical model that describes the dynamics of \textit{B.\,fusca} and provide the mathematical analysis of the model. Section~\ref{sec:biocontrol} discusses biocontrol along with its mathematical analysis. Moreover, this section illustrates the impact of biocontrol on the pest under different cultural practices. Finally, Section~\ref{sec:conclusion} concludes the paper and outlines future directions.

\section{The multi-seasonal model}\label{sec:model-math}
    This section is devoted to the construction, reduction and analysis of a model that describes the epidemiological dynamics within a single plantation during several seasons.

    \subsection{The global model}
    In order to model the infestation by the pest \textit{B.\,fusca}, we first consider the period $T$ which starts from the previous sowing to the next one and $D$ the duration of the cropping season. We divide each period into three phases that represent impulses. Those phases correspond to the emergence phase ($d$ days), the growth-to-harvest phase ($\tau = D-d$ days), and the plant absence phase ($T-D$ days). The latter phase varies significantly: in subsistence farming environments, it lasts longer due to crop rotation methods, whereas in monoculture agricultural systems, it lasts much less. We assume that maize plants are separated into two states: susceptible ones (denoted by $S(t)$) that represents those who never had infectious contacts with pest, and infected ones (denoted by $I(t)$) that was infected by pests. We only consider the stage of pest \textit{B.\,fusca} where they might damage maize plants. Then, we denote by $B(t)$ the population size of \textit{B.\,fusca} at time $t\geq 0$. 
    \medskip
        
\begin{figure}[H]
        \centering
        \includegraphics[width=\textwidth]{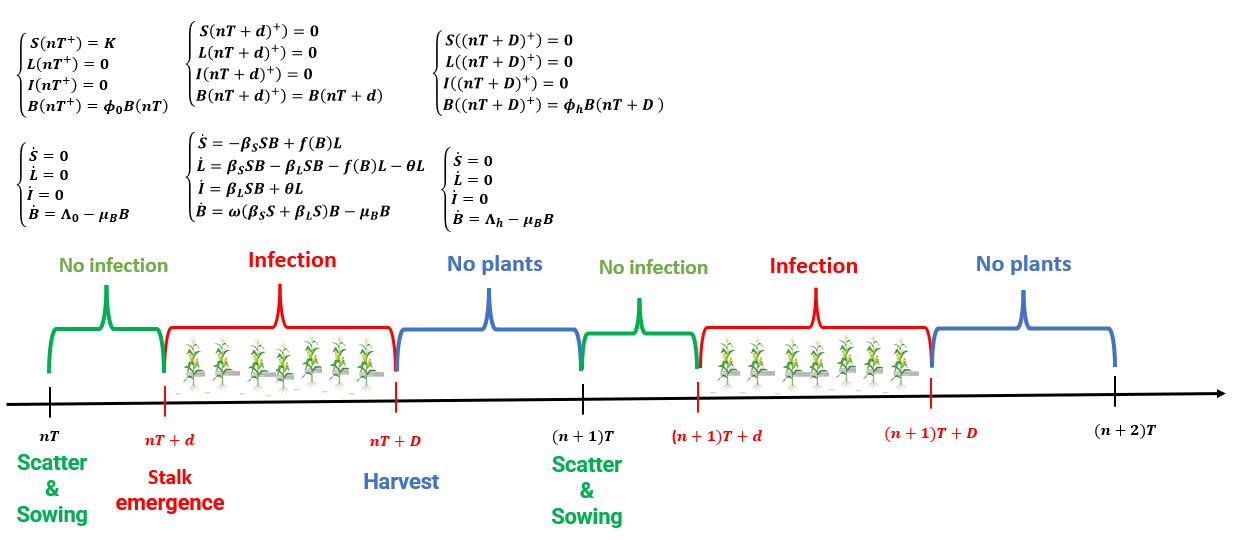}
        \caption{Flochart of dynamics of the \textit{B.\,fusca} with a multi-seasonal model with three impulses. Model equations are given in Eqs.\,\eqref{no_infection}-\eqref{infection_Model}-\eqref{no_plant}.}
        \label{fig:flocahrt-model}
    \end{figure}

    We investigate the case where we are in maize production using the rotation method, and the duration of period $T$ corresponds to the entire year, as is the case in Cameroon for large and small maize farmers \cite{fusillier1993filiere}. We construct a model for the $n$-th year of study, that is the time interval $(nT, (n+1)T]$, and denote by $K$ the total number of maize stalks in the plantation under consideration. The models of each phase are represented in the flowchart given in Figure~\ref{fig:flocahrt-model}. 
    
To build the systems in Figure~\eqref{fig:flocahrt-model}, we will go through three phases: Germination period, Maize production period, and Off-Season.

    \subsubsection{Phase 1 -- Germination period}
The germination phase represents the period from sowing to the emergence of the stalk. According to Harris \& Nwanze \cite{harris1992busseola}, this period lasts between 14 and 30 days, denoted by \(d\).  
At the beginning of the \(n\)-th year, the farmer weeds the plantation before sowing the maize seeds. This cultural practice reduces the pest population remaining from the previous year to a proportion \(\phi_0 \in (0,1)\).  
During this phase, seeds remain in the soil, and we assume they are not susceptible to pest infection. Consequently, the number of susceptible and infected plants remains \(K\) and 0, respectively.  
Additionally, we assume that pests can grow without the stalks at a recruitment rate \(\Lambda_0\), due to immigration or the presence of an alternative host. The pests have a natural mortality rate \(\mu_B\).  We deduce the following systems, so that for $ t \in (nT, nT+d]$ one has:
    \begin{equation}\label{no_infection}
     \left\{
        \begin{array}{ll}
        S(nT^+) =K,\\
         I(nT^+) = 0,\\
         B(nT^+) =\phi_0 B(nT), 
        \end{array}
        \right. \;
        \left\{
    \begin{array}{ll}
      \dot S =0,\\
        \dot  I= 0,\\
        \dot  B =\Lambda_0-\mu_B B
    \end{array}
    \right.
    \end{equation}
    %
    where $S(t)$, $I(t)$ and $B(t)$ represent the number of susceptible plants, infected plants and parasites respectively, at time $t$, so that $S(t) + I(t) = K$.

    \subsubsection{Phase 2 -- Maize production period }
   The maize production phase corresponds to the period from emergence to harvest, lasting \(\tau = D - d\) days. It starts with initial conditions given by the final solution of phase 1. A pest infects a susceptible stalk at a rate \(\beta\). Infected plants can recover at a rate \(f(B)\) and become susceptible again. Biologically, it is reasonable to assume that \(f(B)\) is decreasing, emphasizing that as the number of pests increases, more of them bite the stalks, making recovery more difficult. Therefore, for \(t \in (nT + d, nT + D]\), we obtain the following system:
    
    \begin{equation}\label{infection_Model}
     \left\{
        \begin{array}{ll}
        S((nT+d)^+) =K,\\
         I((nT+d)^+) = I(nT+d),\\
         B((nT+d)^+) = B(nT+d),
        \end{array}
        \right. 
     \left\{
        \begin{aligned}
        \dot S &=-\beta BS+f(B) I,  \\
        \dot I & = \beta BS-f(B) I,\\
        \dot B &=\omega \beta  S B-\mu_B B.
        \end{aligned}
        \right.\;
    \end{equation}

    \subsubsection{Phase 3 -- Off-season }
During this off-season, which lasts \(T - D\) days, alternative crops may be cultivated in a practice known as crop rotation. In countries like Cameroon, fields may be left fallow or planted with beans during this period \cite{fusillier1993filiere}. However, when crop rotation is not practiced, this phase may be too short, leading farmers to start the new maize season by directly planting seeds.  
During this phase, there may or may not be another crop, but the pest population behaves similarly to phase 1, seeking alternative hosts or food sources for survival and reproduction.  
At the end of phase 2, maize plants are removed and seeds are harvested, marking an impulse event at time \(nT + D\). This harvest reduces the pest population to a proportion \(\phi_h \in (0,1)\). Therefore, for \(t \in (nT + D, (n+1)T]\), the following dynamics apply:  
    \begin{equation}\label{no_plant}
     \left\{
        \begin{array}{ll}
        S((nT+D)^+) =0,\\
        I((nT+D)^+) = 0,\\
       B((nT+D)^+) =\phi_h B(nT+D),
        \end{array}
        \right. ~
      \left\{
        \begin{array}{ll}
       \dot S =0,\\
        \dot  I= 0,\\
        \dot  B =\Lambda_h-\mu_B B.
        \end{array}
        \right.\;
    \end{equation}
    
    Table~\ref{tab:param_values} presents the significations of the parameters of models~\eqref{no_infection}-\eqref{infection_Model}-\eqref{no_plant}, their units and values.

    \begin{table}[H]
    \centering
    \caption{ Description and values of parameters for systems~\eqref{no_infection}-\eqref{infection_Model}-\eqref{no_plant}}
    \begin{tabular}{l|p{6cm}|l|l}
     &  Biological meaning & Literature Values  & Value \\\hline
    $d$& Duration of the germination  & $[14, 30]$ days \cite{harris1992busseola} & $20$\\
    $D$&  Duration of the Cropping season & $[90,130]$ days \cite{hoopen2012production} &$120$\\
    $K$ & Number of plant per hectare & $[53, 70]\times 10^3$ plant.ha$^{-1}$ \cite{ngoune2019estimation} & $60000$ \\
    $\beta$ & Infestation rate & $[0.11 ,0.67]/K$ (pests.day)$^{-1}$ \cite{taddele2020spatial}&$0.3/K$\\
    $\omega$ & Number of pest generate by a single infection & $[0,5]$ pests.plant$^{-1}$ \cite{usua1968effect}&$2$\\
    $\Lambda_0$ &  Constant recruitment of pest & $[4,116]$ pests.day$^{-1}$ \cite{Calatayud2007}&$5$\\
    $\Lambda_h$ &  Constant recruitment of pest & $[4,116]$ pests.day$^{-1}$ \cite{Calatayud2007}&$8$\\
    $\mu_B$ & Mortality rate of  pests &  $0.35$ day$^{-1}$ \cite{ntahomvukiye2018study}&$0.35$\\
    $\phi_0$ & Fraction of pest destroyed by sacking process & /day$^{-1}$& 0.8\\
    $\phi_h$ & Fraction of pest destroyed by harvesting process & /day$^{-1}$& 0.5\\
    \hline
    \end{tabular}
    \label{tab:param_values}
    \end{table}

    \subsection{Reduction of the model}
In the systems~\eqref{no_infection}–\eqref{infection_Model}–\eqref{no_plant}, since there is no infection in the interval \((nT, nT + d]\), we can solve the system, determine the expression for the pest population, and consider the beginning of cropping at \(nT + d\). Moreover, since the total number of maize plants remains constant, we only consider the system for \(I\) and \(B\), as \(S = K - I\).

    Then for the mathematical analysis of \eqref{no_infection}-\eqref{infection_Model}-\eqref{no_plant}, we consider a reduced model $(I_1,B_1)$ that consists of taking the dynamics of $(I,B)$ only on the intervals $\cup_{n\geq 0}[nT+d,nT+D)$, that is for the corresponding time $t\in(n\tau, (n+1)\tau]$. We determine the expressions for the impulses of \( B \) by explicitly solving the equations \( \dot{B} = \Lambda_0 - \mu_B B \) and \( \dot{B} = \Lambda_h - \mu_B B \), in order to find the solution of the model during the germination period and the off-season, respectively. Finally, the reduced system is given by:
    
    \begin{equation}\label{reduce_multi_Model}
        \left\{
        \begin{aligned}
         &  I_1(n\tau^+) = 0,\\
           &  B_1(n\tau^+) =\phi_0\left[\frac{\Lambda_0}{\mu_B}+ \left[\frac{\Lambda_h-\Lambda_0}{\mu_B} + \left(\phi_h B_1(n\tau)-\frac{\Lambda_h}{\mu_B}\right)e^{-\mu_B(T-D)}\right]e^{-\mu_Bd}\right],\\
           & \dot I_1  = \beta B_1(K-I_1) -f(B_1) I_1\\
            & \dot B_1 =\omega \beta (K-I_1)B_1-\mu_B B_1.
        \end{aligned}
        \right.
    \end{equation}


    \subsection{Mathematical analysis}
  
   This section is devoted to the mathematical analysis of the reduced model~\eqref{reduce_multi_Model}. We focus primarily on the stability of the pest-free solution when maize is the only resource available to pests, i.e., \(\Lambda_0 = \Lambda_h = 0\). Thus, we derive conditions that ensure the extinction of pests in the plantation.

    \subsubsection{Periodic pest-free solution and its stability}
   
    When $\Lambda_0=\Lambda_h=0$, the reduced model given by Eq.\,\eqref{reduce_multi_Model} admits a unique periodic pest free solution (PPFS) given by $X^T(t) = (0,0)$ for any $t\geq 0$. We deduce the main result of this section that follows.
    \begin{theorem}\label{theo:gas}
        Let us introduce the following parameters
        \begin{equation}
            \mathcal{R}_0 = \frac{\beta\omega K}{\mu_B} ~\textrm{ and }~ \mathcal{R} = \phi_0\phi_h e^{-\mu_B\left( T - \mathcal{R}_0\tau \right)},
        \end{equation}
        then the PPFS $X^T(t)$ is locally asymptotically stable if $\mathcal{R}<1$ and unstable otherwise. Moreover, it is globally asymptotically stable if $\mathcal{R}_0<1$.
    \end{theorem}
    \medskip

    \begin{proof}
        The proof of the local stability is based on the linearization approach using Floquet Theory \cite{bainov2017impulsive}. The linearization of system \eqref{reduce_multi_Model} around the PPFS $X^T(t)$ gives
        \begin{equation}\label{linear_Model}
            \left\{
            \begin{aligned}
           & \dot {\tilde X}(t) =A {\tilde X}(t), \; t\neq n\tau\\
            & \dot {\tilde X}(n\tau^+)=\mathrm{diag}(0,\phi_B) {\tilde X}(n\tau)
            \end{aligned}
            \right.
        \end{equation}
        where $\phi_B=\phi_0 \phi_h e^{-\mu_B(T-\tau)}$ and 
        $$A=\begin{pmatrix}-f(0)  & \beta K\\ \\
        0& \omega \beta K-\mu_B\end{pmatrix} .$$
        Solving the first equation of~\eqref{linear_Model} with $t\in(0,\tau]$, we obtain 
        $$\tilde X(t)=\Phi_A(t)X(0^+)$$
        where
        $$\Phi_A(t)=e^{At}=\begin{pmatrix} e^{-f(0)t}  & *\\ \\
        0& e^{(\omega \beta K -\mu_B)t} \end{pmatrix} $$
        is the fundamental matrix. Using the second  equation of~\eqref{linear_Model},
        we obtain 
        $$ \tilde X((n+1)\tau^+)= M X(n \tau ^+)$$
        where $M= \mathrm{diag}(0,\phi_B) e^{A\tau}$ is the monodromy matrix of~\eqref{linear_Model}. Replacing the expression of the exponential matrix, we obtain:
        \[
        M=\begin{pmatrix}
        0&*\\
        0& \mathcal{R} \end{pmatrix} .
        \]
        Due to the block-triangular form of monodromy matrix $M$, there is no need to calculate the exact form of $(*)$ for the following analysis.  Floquet multipliers of $M$ are given by $0$ and 
        \[
        \mathcal{R}=\phi_0 \phi_h e^{-\mu_BT+\omega \beta K\tau}.
        \]
        The  PPFS is asymptotically stable if the Floquet multipliers of the monodromy matrix $M$ belongs to the unit circle, i.e $\mathcal{R}<1$, and unstable otherwise.

        Further, if we denote by $y(t) = (y_1(t),y_2(t)) = (I_1(t),B_1(t))$ for any $t\geq 0$, then the reduced model \eqref{reduce_multi_Model} can be rewritten as follows
        \begin{equation}\label{sub_stability}
            \left\{
            \begin{aligned}
                & \dot y(t) = F_c(y(t)), t\neq n\tau \\
                &  \Delta y(t) =F_d(y(t)), t=n\tau\\
            \end{aligned}
            \right.
        \end{equation}
        where 
        $F_c(y)=\begin{pmatrix} 
        \beta (K-y_1)y_2 -f(y_2)y_1 \\ 
        \omega \beta(K-y_1)y_2  -\mu_B y_2
        \end{pmatrix}$ and 
        $F_d(y)=\begin{pmatrix} 
        -y_1\\
        ( \phi_B-1)y_2\end{pmatrix}$. \\

        Let us consider the Lyapunov function candidate \( V: \mathbb{R}^2_+ \to [0, \infty) \) defined by  
\[
V(y_1, y_2) = y_1 + a y_2
\]
where \( a \) is a non-negative constant to be determined later.
        \begin{itemize}
            \item[(i)]   The above function satisfies $V(0,0)=0$, $V(y)>0$ for any $y\in  \mathbb{R}^2_+-\{(0,0)\}$ and $V(y)\to +\infty$ as $\|y\| \to +\infty$.
            \item [(ii)] Let us prove that $ V'(y)F_c(y)\leq 0$
            \begin{align*}
                V'(y)F_c(y)&= \beta (K-y_1)y_2 -f(y_2)y_1+a(\omega \beta(K-y_1)  -\mu_B) y_2\\
                & =  -f(y_2) y_1-\beta(1+\omega) y_1 y_2+ (\beta K + a(\omega \beta K-\mu_B))y_2 .
            \end{align*}
            We obtain $ V'(y)F_c(y) \leq  -f(y_2) y_1-\beta(1+\omega) y_1 y_2$ if $(\beta K + a(\omega \beta K-\mu_B))=0$, that is
            $$  a=\frac{\beta K}{\mu_B-\omega \beta K} $$ 
            which is well defined if and only if $\mu_B-\omega \beta K > 0$, i.e $\mathcal{R}_0= \frac{\beta \omega K}{\mu_B} < 1$.
        
            \item [(iii)] We also have 
            \begin{align*}
                V(y+F_d(y))&=y_1-y_1+a(y_2+( \phi_B-1)y_2)\\
                & \leq y_1+a y_2\\
                & \leq V(y) .
            \end{align*}
        
        \end{itemize}
        Using \cite[Theorem 4.1]{haddad2006impulsive},  we deduce thanks to (i)--(iii) here above that the PPFS $X^T(t)$ is globally asymptotically stable under the condition $\mathcal{R}_0 < 1$.
        
    \end{proof}

    When $\Lambda_0\neq 0$ or $\Lambda_h \neq 0$, the PPFS does not exist because pests are consistently present in the plantation. In some cases, this quantity may be very low, which can lead to insufficient infection levels. Let us perform numerical simulations of the model in order to illustrate its behaviour. 

    \subsection{Numerical simulation}
    We consider the initial systems~\eqref{no_infection}-\eqref{infection_Model}-\eqref{no_plant} where we add the dynamics of susceptible plants. We do not represent the germination and off-season periods. The function $f$ is defined using the approach developed by~\cite{anguelov2012mathematical} from which it has the following form:  
    \begin{equation}\label{eq:funtion_f}
        f(B)=\eta e^{-\frac{\beta}{\eta} B}
    \end{equation}
    where $1/\eta$ is the average time of existence of the infestation 
    on the infected plant and $\beta$ the infestation rate. The value of $\beta$ is given in Table~\ref{tab:param_values}  and we suppose that $\eta=0.05$ day$^{-1}$.
    
    We opted to simulate the dynamics in an area of \( 50 \, \text{m}^2 \) with a total of \( K = 300 \) plants, which is equivalent to $60000$ plants.ha$^{-1}$ considered from the literature\cite{cropnuts_maize_population,seedco_sc719,TSAFACK2024e35660}. The initial conditions are set as follows:
    \begin{equation}\label{eq:initial_values}
        (S(0^+),I(0^+),B(0^+)) = (K,0,100)
    \end{equation}
    
\begin{figure}[H]
        \centering
        \begin{subfigure}[b]{0.45\textwidth}
            \centering
            \caption{}
            \includegraphics[width=\textwidth]{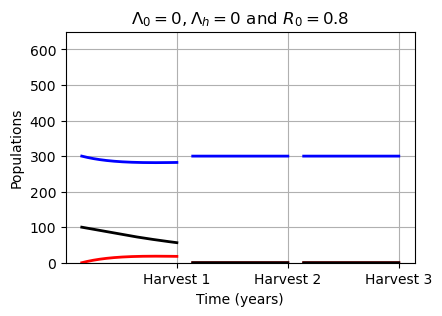}
            \label{fig:lambda0R0.8}
        \end{subfigure}
        \begin{subfigure}[b]{0.45\textwidth}
            \centering
             \caption{}
            \includegraphics[width=\textwidth]{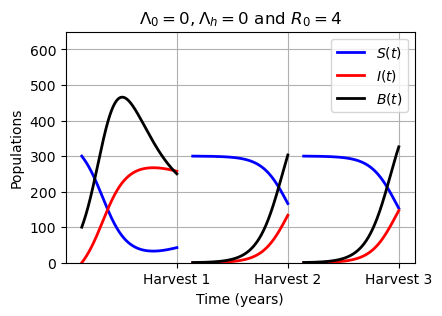}
            \label{fig:lambda0R04}
        \end{subfigure}
        
        \vspace{-0.5cm}
        \begin{subfigure}[b]{0.45\textwidth}
            \centering
            \caption{}
            \includegraphics[width=\textwidth]{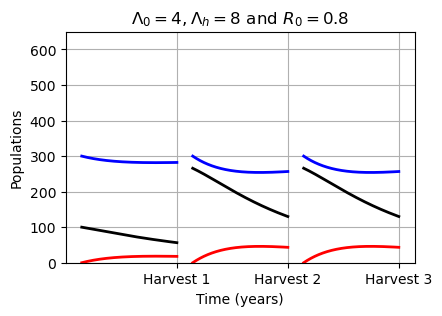}
            \label{fig:lambdanonnulR0.8}
        \end{subfigure}
        \begin{subfigure}[b]{0.45\textwidth}
            \centering
            \caption{}
            \includegraphics[width=\textwidth]{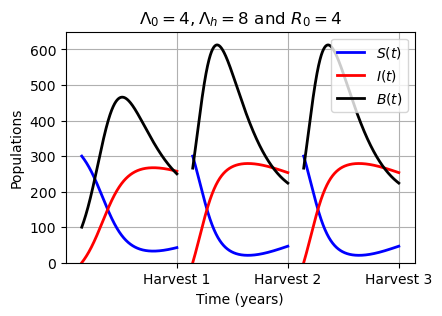}
            \label{fig:lambdanonnulR04}
        \end{subfigure}
        \vspace{-0.7cm}
        \caption{Impact of pests on the dynamics of maize plants. The plots present the trajectories of~\eqref{no_infection}-\eqref{infection_Model}-\eqref{no_plant} when the PPFS is stable (Subplot~\ref{fig:lambda0R0.8}), $\mathcal{R}_0=0.8 \leq1$, and when the PPFS is unstable (subplot~\ref{fig:lambda0R04}), $\mathcal{R}_0=4 >1$. Subplots~\ref{fig:lambdanonnulR0.8} and \ref{fig:lambdanonnulR04} represent the dynamics for $\Lambda_0=4$ and $\Lambda_h=8$ when $\mathcal{R}_0=0.8$ and $\mathcal{R}_0=4$, respectively. The remaining parameter values are given in Table~\ref{tab:param_values}, and initial conditions are provided by Eq.~\ref{eq:initial_values}.}
        \label{fig:ide_reduce_model}
    \end{figure}

    Figure~\ref{fig:ide_reduce_model} illustrates the dynamics of~\eqref{no_infection}-\eqref{infection_Model}-\eqref{no_plant}. Subplots~\ref{fig:lambda0R0.8} and \ref{fig:lambda0R04} depict the cases where the constant recruitment of pests during the germination period (\( \Lambda_0 \)) and the off-season (\( \Lambda_h \)) are both set to zero. In Subplot~\ref{fig:lambda0R0.8}, the extinction of pests occurs when \( \mathcal{R}_0 = 0.8 \leq 1 \), while Subplot~\ref{fig:lambda0R04} demonstrates the persistence of pests when \( \mathcal{R}_0 = 4 > 1 \), confirming the results presented in Theorem~\ref{theo:gas}.
    
    For the case where \( \Lambda_0 = 4 \) and \( \Lambda_h = 8 \), 
    Subplot~\ref{fig:lambdanonnulR0.8} illustrates that pests persist in the plantation even when the basic reproduction number is below one, meaning \( \mathcal{R}_0 = 0.8 \leq 1 \).  
Furthermore, in the scenario where \( \Lambda_0 \) and \( \Lambda_h \) are nonzero, Subplot~\ref{fig:lambdanonnulR04} shows that with \( \mathcal{R}_0 = 4 > 1 \), pest infestations increase progressively over each period. This contrasts with the case where both \( \Lambda_0 \) and \( \Lambda_h \) are zero, as seen in Subplot~\ref{fig:lambda0R04}, where pest infestation peaks only once during the first year.  
These subplots also highlight that by the end of the third period, after the harvest of year three, the number of susceptible plants remains at 100 when \( \Lambda_0 \) and \( \Lambda_h \) are both zero. However, when these values are nonzero, the number of susceptible plants drops below 50.  This result demonstrates the importance of analyzing pest dynamics during the germination period and off-season to develop more effective pest biocontrol strategies.

\section{Biocontrol of pest using predator}\label{sec:biocontrol}

We consider the global multi-seasonal model presented in Section~\ref{sec:model-math}, which captures the dynamics of maize growth and pest interactions across multiple seasons.
  The model that the maize growth cycle into distinct phases: germination, cropping season, and off-season. The model tracks the populations of susceptible plants (\(S\)), infected plants (\(I\)), and pests (\(B\)), with their interactions described by impulsive differential equations. Additionally, we apply biocontrol measures by releasing predators that consume \textit{B. fusca}.

    \subsection{The yearly quantity and release number are constant }
Based on \cite{Djuikem_impulsive}, we assume that the yearly quantity \(\Lambda_P\) of predators is released at multiple times throughout the year. This quantity is evenly distributed into \(m\) releases, each of size \(\frac{\Lambda_P}{m}\), with a release interval of \(\frac{\tau}{m}\).  
In contrast to \cite{Djuikem_impulsive}, where predators are released at the beginning of the year, in our model, they are released only after \(d\) days. Moreover, as predators of prey \(B\), the predator population \(P\) is present throughout all phases and consumes the prey according to a Holling type 2 functional response. This leads to an impulsive model with a modified switching condition for the predator, described by the following system.
    
     For $ t \in (nT, nT+d]$
    
    \begin{equation}\label{no-inf-bioBP}
        \left\{
        \begin{array}{ll}
        S(nT^+) =0, \\
        I(nT^+) =0, \\
         B(nT^+) =\phi_0 B(nT), \\
        P(nT^+) =\phi_0 P(nT),\\
         \begin{array}{ll}
        \dot  S =K,\\
        \dot  I =0,\\
        \dot  B =\Lambda_0-\mu_B B-\alpha \frac{BP}{B+K_B} ,\\
      \dot   P = \nu \alpha \frac{BP}{B+K_B}-\mu_P P.
      \end{array} 
    \end{array}
    \right.
    \end{equation}
    
    For $t\in(nT+d+\frac{j\tau}{m}, nT+d+\frac{(j+1)\tau}{m}]$:
    \begin{equation}\label{inf-bioBP}
     \left\{
        \begin{array}{ll}
         S((nT+d)^+) = K, \\
         I((nT+d)^+) = I(nT+d), \\
         B((nT+d)^+) = B(nT+d), \\
       P((nT+d+\frac{j\tau}{m})^+)=P(nT+d+\frac{j\tau}{m})+\frac{\Lambda_P}{m},\; j \in \{0,...,m-1\} \\
       \begin{array}{ll}
        \dot S = -\beta BS+f(B) I, \\
        \dot I = \beta BS-f(B) I, \\
        \dot B =\omega \beta SB-\mu_B B-\alpha \frac{BP}{B+K_B} ,\\
     \dot  P = \nu \alpha \frac{BP}{B+K_B}-\mu_P P.
        \end{array}
        \end{array}
        \right.
    \end{equation}

    For \( t \in (nT+D, (n+1)T] \):
    \begin{equation}\label{no-plant-bioBP}
        \left\{
        \begin{array}{ll}
            S((nT+D)^+) = 0,\\
             I((nT+D)^+) = 0,\\
             B((nT+D)^+) = \phi_h B(nT+D),\\
             P((nT+D)^+) = \phi_h P(nT+D),\\
            \begin{array}{ll}
              \dot{S} = 0,\\
             \dot{I} = 0,\\
             \dot{B} = \Lambda_h - \mu_B B - \alpha \frac{BP}{B+K_B} , \\
             \dot{P} = \nu \alpha \frac{BP}{B+K_B} - \mu_P P.
            \end{array} 
        \end{array}
        \right.
    \end{equation}

For the mathematical analysis of systems~\eqref{no-inf-bioBP}-\eqref{inf-bioBP}-\eqref{no-plant-bioBP}, we remove the equation for susceptible plants and use the fact that the total number of plants remains constant (\( K = S + I \)). Thus, in the other equations, \( S \) is replaced by its equivalent expression \( S = K - I \).

    \subsubsection{Multiple release controlled periodic disease-free solution and its stability }
    We can compute and analyze the multiple release controlled periodic disease-free solution (m-cPDFS) in the case where $\Lambda_0=\Lambda_h=0$, 

    For $ t \in (nT, (n+1)T]$
    \begin{equation}\label{eq_many_predator}
        \left\{
        \begin{aligned}
            & \dot   P = -\mu_P P,\; t\neq nT, t\neq nT+d+\frac{j\tau}{m},\; j \in \{0,...,m-1\}, \\
            & P(d^+) =P(d),\\
            & P\left(\left(d+\frac{j\tau}{m}\right)^+\right)=P\left(d+\frac{j\tau}{m}\right)+\frac{\Lambda_P}{m},\; j \in \{0,...,m-1\}, \\
            & P(D^+) =\phi_h P(D),\\
            & P(T^+) =\phi_0 P(T).\\
        \end{aligned}
        \right.
    \end{equation}
   Solving system~\eqref{eq_many_predator}, the m-cPDFS of system formed by systems~\eqref{no-inf-bioBP}-\eqref{inf-bioBP}-\eqref{no-plant-bioBP}  is $Y^T(t)=(0,0,P^T(t))$, where the expression of $P^T(t)$ is established in Appendix~\ref{sec:app_m-cPDFS} and given by
    \begin{equation}\label{Multi_Period_sol2}
        P^T(t)=\left\{
        \begin{aligned}
            & P^s(t)=P(0^+)e^{-\mu_P t},\; t\in I_s\\
            & P^c_{m,j}(t)=P\left(0^{+}\right)e^{-\mu_P t}+ \frac{\Lambda_P}{m}\left(\frac{1-e^{-\mu_P \frac{(j+1)\tau}{m}}}{1-e^{-\mu_P \frac{\tau}{m}}}\right) e^{-\mu_P \left(t-d-\frac{j\tau}{m}\right)}, \; t\in I_{c,j} \\
            &P^o(t)=\phi_h\left[P\left(0^{+}\right)e^{-\mu_P \tau}+ \frac{\Lambda_P e^{-\mu_P\frac{\tau}{m}}}{m}\left(\frac{1-e^{-\mu_P\tau}}{1-e^{-\mu_P \frac{\tau}{m}}}\right)\right]e^{-\mu_P (t-D)}, \; t \in I_o.
        \end{aligned}
        \right.
    \end{equation}
    with $I_h=(0,d]$, $I_{c,j}=(d+\frac{j\tau}{m}, d+\frac{(j+1)\tau}{m}]$, $j=1,\ldots,m-1$, $I_o=(D, T]$, and
    \begin{equation}\label{Fixe_pt_Multi}
        P\left(0^{+}\right)=  \frac{ \phi_0 \phi_h\Lambda_P}{m}\left[ \frac{e^{-\mu_P(T-D)}-e^{-\mu_P (T-d)}}{\left(1-\phi_0\phi_h e^{-\mu_P T}\right)\left(e^{\mu_P \frac{\tau}{m}}-1\right)}\right]. 
    \end{equation}

    \begin{lemma}\label{lem_many_release}
        The m-cPDFS $Y^T(t)=(0,0,P^T(t))$ of the controlled model~\eqref{no-inf-bioBP}-\eqref{inf-bioBP}-\eqref{no-plant-bioBP}  with multiple releases, is locally asymptotically stable when $\mathcal{R}_c<1$ and unstable otherwise, where
        \begin{equation}\label{eq_Rc}
            \mathcal{R}_c=\phi_0 \phi_h e^{-\mu_B T+ \omega \beta K\tau -\frac{\alpha}{K_B}\int_{0}^T P^T(a)da}.
        \end{equation}
    \end{lemma}

    \begin{proof}
        For \( t \in (nT, nT + d] \), the linearization of the system~\eqref{no-inf-bioBP} around the m-cPDFS is given by:
        
        \begin{equation}\label{linear_Model_s}
   \dot {\tilde X}(t) =A_s(t) {\tilde X}
        \end{equation}
        where
        $$
        A_s(t)=\begin{pmatrix}
        0 & 0 & 0 \\
        0 & -\mu_B-\frac{ \alpha P^T(t)}{K_B}& 0 \\
        0 & \frac{\nu \alpha P^T(t)}{K_B} &-\mu_P  
        \end{pmatrix} .
        $$

        Solving the first equation of system~\eqref{linear_Model_s} for $t \in (0,d]$, we obtain 
        $$\tilde X(t)=\Phi_{A_s}(t)X(0^+)$$
 where $\Phi_{A_s}(t)$ is the fundamental matrix define by:
        $$
        \Phi_{A_s}(t)=\begin{pmatrix}
        1 & 0 & 0 \\
        0 & e^{-\mu_Bt-\int_{0}^{t}\frac{ \alpha P^T(a)}{K_B}da}& 0 \\
        0 &* &e^{-\mu_P t} 
        \end{pmatrix}
        $$
      Using the impulse at the end of the germination phase $\tilde X((nT+d)^+)= {\tilde X}(nT+d)$, we obtain $ X((nT+d)^+)= M_sX(nT^+)$, where $M_s=\Phi_{A_s}(d)$.

        For $t\in\Big(nT+d+\frac{j\tau}{m}, nT+d+\frac{(j+1)\tau}{m}\Big]$ with $j=0...,m-1$, the linearization of the system~\eqref{inf-bioBP} around the m-cPDFS is given by
        \begin{equation}\label{linear_Model_cj}
            \left\{
            \begin{aligned}
            & \dot {\tilde X}(t) =A_{c,j}(t) {\tilde X}(t), \\
            & \tilde X((nT+d + \frac{j\tau}{m})^+)={\tilde X}(nT+d + \frac{j\tau}{m})
            \end{aligned}
            \right.
        \end{equation}
        where $$A_{c,j}(t)=\begin{pmatrix}-f(0)  & \beta K & 0\\ 
        0& \omega \beta K-\mu_B-\alpha\frac{P^c_{m,j}(t)}{K_B} & 0\\
        0 &  \nu \alpha\frac{P^c_{m,j}(t)}{K_B} & -\mu_P
        \end{pmatrix}.$$
        Recall that $P^c_{m,j}(t)$ is the restriction of $P^T(t)$ on the time interval $I_{c,j} = (d+\frac{j\tau}{m},d+\frac{(j+1)\tau}{m}]$. Then it follows from Eqs.\,\eqref{linear_Model_cj} with $j=0,...,m-1$ that the linearization of \eqref{inf-bioBP} around the m-cPDFS on the whole interval $(nT+d,nT+D]$ is given by
        \begin{equation}\label{linear_Model_c}
            \dot {\tilde X}(t) =A_{c}(t) {\tilde X}(t), \forall t\in(nT+d,nT+D]
        \end{equation}
        where $$A_{c}(t)=\begin{pmatrix}-f(0)  & \beta K & 0\\ 
        0& \omega \beta K-\mu_B-\alpha\frac{P^T(t)}{K_B} & 0\\
        0 &  \nu \alpha\frac{P^T(t)}{K_B} & -\mu_P
        \end{pmatrix}.$$
        Solving Eq.\,\eqref{linear_Model_c} on $(d,D]$, we obtain:
        $$\tilde X(t)=\Phi_{A_c}(t)X(d^+)$$
        where
        $$\Phi_{A_c}(t)=\begin{pmatrix} e^{-f(0) (t-d)}  &* & 0\\ 
        0& e^{(\omega \beta K-\mu_B)(t-d)-\frac{\alpha}{K_B}\int_{d}^{t} P^T(a)da}& 0\\
        0 &  * & e^{-\mu_P(t-d)}
        \end{pmatrix} .$$
        In addition, the impulse at the end of the maize production period is given by $$X((nT+D)^+)= diag (0,\phi_h, \phi_h)X(nT+D).$$ It follows that $ X((nT+D)^+)= M_cX((nT+d)^+)$, where $M_c=diag(0,\phi_h,\phi_h)\Phi_{A_c}(D)$. 

        For \( t \in (nT+D, (n+1)T] \), the linearization of the system~\eqref{inf-bioBP}  around the m-cPDFS is given by
        \begin{equation}\label{linear_Model_o}
            \left\{
            \begin{aligned}
            & \dot {\tilde X}(t) =A_o(t) {\tilde X}(t), \; t\neq nT+D\\
            & \tilde X((n+1)T^+)=diag(0,\phi_0, \phi_0) {\tilde X}((n+1)T)
            \end{aligned}
            \right.
        \end{equation}
        where
        $$
         A_o(t)=\begin{pmatrix}
         0 & 0 & 0 \\
         0 & -\mu_B-\frac{ \alpha P^T(t)}{K_B}& 0 \\
         0 & \frac{\nu \alpha P^T(t)}{K_B} &-\mu_P  
        \end{pmatrix} .
        $$
        Solving the first equation of system~\eqref{linear_Model_o} for $t \in (D,T]$, we obtain:
        $$\tilde X(t)=\Phi_{A_o}(t)X(D^+)$$
        where 
        $$
        \Phi_{A_o}(t)=\begin{pmatrix}
        1& 0 & 0 \\
        0 & e^{-\mu_B(t-D)-\frac{\alpha}{K_B}\int_{D}^tP^T(a) da}& 0 \\
        0 & *&e^{-\mu_P(t-D)}  
        \end{pmatrix} .
        $$
        Using the impulse in the second  equation of system~\eqref{linear_Model_o}, we obtain
        $ X((n+1)T^+)= M_oX((nT+D)^+)$ where $M_o= diag(0,\phi_0, \phi_0)\Phi_{A_o}(T)$. 

        We deduce from the computations here above that for $t\in (nT, (n+1)T]$, the solution of the controlled systems~\eqref{no-inf-bioBP}-\eqref{inf-bioBP}-\eqref{no-plant-bioBP} satisfies 
        $$\tilde X((n+1)T^+)= \mathcal{M} X(nT^+)$$
        where $\mathcal{M}=M_o M_c M_s$ is the monodromy matrix. 
        Replacing $M_o$, $M_c$ and $ M_s$ by their expressions, we obtain
        $$\mathcal{M}=M_o M_c M_s= \begin{pmatrix}
         0& 0 & 0 \\
         0 & \mathcal{R}_c& 0 \\
         0 & *&\phi_0 \phi_h e^{-\mu_p T}
        \end{pmatrix} $$
        where $\mathcal{R}_c=\phi_0 \phi_h e^{-\mu_B T+ \omega \beta K\tau-\frac{\alpha}{K_B}\int_{0}^T P^T(a)da}$.
       
        Then the m-cPDFS is locally asymptotically stable if and only if $\mathcal{R}_c<1$.
        
    \end{proof}
Compared to the model without biocontrol, the release of predators contributes to the stabilization of the PDFS, as seen in the formula  

\[
\mathcal{R}_c = \mathcal{R}e^{-\frac{\alpha}{K_B} \int_0^T P^T(a) \, da}.
\]  
This shows that the total number of predators over time, \(\int_0^T P(a) \, da\), exponentially reduces the threshold \(\mathcal{R}\) obtained in the model without biocontrol.

    \subsection{Impact of biocontrol and cultural practice on the pest dynamics}
   In this section, in addition to biocontrol, we consider cultural practices as an important control measure for smallholder farmers \cite{fusillier1993filiere}. We simulate different scenarios to assess the effectiveness of combining biocontrol with cultural practices.  

\begin{figure}[H]
        \centering
        \begin{subfigure}[b]{0.45\textwidth}
            \centering
            \caption{}
            \includegraphics[width=\textwidth]{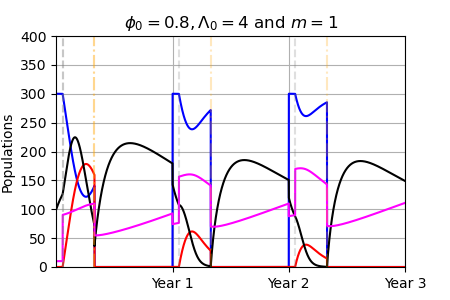}
            \label{fig:biocontrol_m1_phi_0.8}
        \end{subfigure}
        \begin{subfigure}[b]{0.45\textwidth}
            \centering
             \caption{}
            \includegraphics[width=\textwidth]{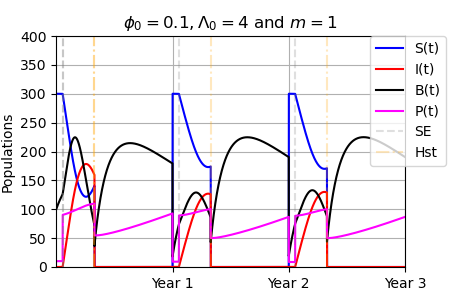}
            \label{fig:biocontrol_m1_phi_0.1}
        \end{subfigure}
        
        \vspace{-1cm}
        \begin{subfigure}[b]{0.45\textwidth}
            \centering
            \caption{}
            \includegraphics[width=\textwidth]{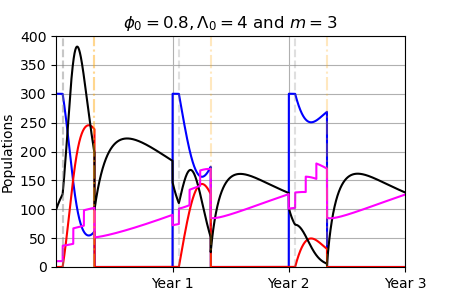}
            \label{fig:biocontrol_m3_phi_0.8}
        \end{subfigure}
        \begin{subfigure}[b]{0.45\textwidth}
            \centering
            \caption{}
            \includegraphics[width=\textwidth]{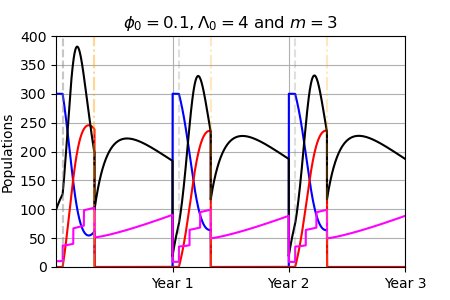}
            \label{fig:biocontrol_m3_phi_0.1}
        \end{subfigure}
        \vspace{-0.7cm}
        \caption{ Effect of biocontrol and cultural practices on pest dynamics in a maize plantation. The plots show the trajectories of the global impulsive biocontrol systems~\eqref{no-inf-bioBP}-\eqref{inf-bioBP}-\eqref{no-plant-bioBP} with \(\mathcal{R}_0 = 4\) and \(\Lambda_0 = 4\). Subplots~\ref{fig:biocontrol_m1_phi_0.8} and \ref{fig:biocontrol_m1_phi_0.1} represent the dynamics for a single predator release per year (\(m=1\)) under good (\(\phi_0 = 0.8\)) and poor (\(\phi_0 = 0.1\)) cultural practices, respectively. Subplots~\ref{fig:biocontrol_m3_phi_0.8} and \ref{fig:biocontrol_m3_phi_0.1} illustrate the dynamics when predators are released three times per year (\(m=3\)) under good and poor cultural practices. The remaining parameter values are provided in Table~\ref{tab:param_values}, and initial conditions are given by Eq.~\eqref{eq:initial_values}.}
        \label{fig:biocontrol_Cas1}
    \end{figure}

Figure~\ref{fig:biocontrol_Cas1} illustrates the dynamics of susceptible maize plants $S$ (blue curve), infected plants  $I$(red curve), pests $B$ (black curve), and pest predators $P$ (magenta curve) for a basic reproduction number of \(\mathcal{R}_0 = 4\) and pest recruitment during the germination period with \(\Lambda_0 = 4\) and \(\Lambda_h = 8\). The initial conditions are defined by equation~\eqref{eq:initial_values}, with an initial predator population of \( P_0 = 10 \).  
The gray dashed line indicates the moment of stalk emergence (SE), while the orange dashed line marks the time of harvest (Hst). A fixed quantity of predators, \(\Lambda_P = 80\), is released either once (\( m = 1 \)) or three times (\( m = 3 \)) per year. The cultural practice is classified as good or poor, corresponding to the values \(\phi_0 = 0.8\) and \(\phi_0 = 0.1\), respectively.

Figures~\ref{fig:biocontrol_m1_phi_0.8} and \ref{fig:biocontrol_m1_phi_0.1} show the dynamics of systems~\eqref{no-inf-bioBP}-\eqref{inf-bioBP}-\eqref{no-plant-bioBP} when predators are released once per year (\(m=1\)), under good and poor cultural practices, respectively.  
After the first year, the number of susceptible plants stabilizes at around 140 for both good and poor cultural practices. However, in the second and third years, Figure~\ref{fig:biocontrol_m1_phi_0.8} shows that with biocontrol and good cultural practices, the number of susceptible plants increases to approximately 250 in the second year and 275 in the third year. In contrast, Figure~\ref{fig:biocontrol_m1_phi_0.1} demonstrates that under poor cultural practices, the number of susceptible plants remains at 175 in both the second and third years. This indicates that, with biocontrol and poor cultural practices, there are approximately 100 more infected plants compared to the scenario with good cultural practices.  

Figures~\ref{fig:biocontrol_m3_phi_0.8} and \ref{fig:biocontrol_m3_phi_0.1} present the dynamics of systems~\eqref{no-inf-bioBP}-\eqref{inf-bioBP}-\eqref{no-plant-bioBP} when predators are released three times per year (\(m=3\)), under good and poor cultural practices, respectively. After the first year, the number of susceptible plants stabilizes at approximately 50 for both scenarios. However, Figure~\ref{fig:biocontrol_m3_phi_0.8} shows that with biocontrol and good cultural practices, the number of susceptible plants increases to about 175 in the second year and 270 in the third year. In contrast, Figure~\ref{fig:biocontrol_m3_phi_0.1} shows that under poor cultural practices, the number of susceptible plants remains close to 70 in both the second and third years, resulting in approximately 200 more infected plants compared to the case with good cultural practices.  

Figure~\ref{fig:biocontrol_Cas1} shows that under good cultural practices, the number of susceptible plants in the third year is nearly identical for both \(m=1\) and \(m=3\) release strategies. This suggests that, asymptotically, both strategies provide farmers with the most effective long-term pest control. However, in the case of poor cultural practices, biocontrol is significantly less effective. Specifically, for \(m=3\), the number of infected plants is almost the same as in the scenario without control, highlighting the critical role of cultural practices.  
In conclusion, the most effective strategy appears to be a single predator release per year (\(m=1\)), regardless of whether cultural practices are good or poor.
    
\section{Conclusion}\label{sec:conclusion}

This study presented the mathematical modeling and biological control of \textit{B.\,fusca} using a multiseasonal model. Using Floquet theory, we analyzed the stability of periodic solutions and computed the conditions for pest extinction in the plantation when there is no alternative resource and no immigration. Numerical simulations were performed to observe the impact of different factors, including predator release frequencies and environmental carrying capacity. The results demonstrated that biological control could serve as a sustainable alternative to chemical pesticides. However, when combined with good cultural practices, this biocontrol becomes more efficient in reducing the population of \textit{B.\,fusca} over multiple growing seasons.
Future studies could refine the model by considering additional ecological factors, such as climate variability and crop rotation effects, to further improve pest control strategies.

\newpage
\appendix

    \section{Expression of the m-cPDFS}\label{sec:app_m-cPDFS}
  Let us solve system~\eqref{eq_many_predator} over the interval \( (0, T] \). Solving equation $\dot{P} = -\mu_P P$ for $ t\in (0,d]$, we obtain
    \begin{equation}\label{Multi_Period_sol_s}
        P^s(t)=P(0^+)e^{-\mu_Pt}
    \end{equation}
    and for $ t\in I_{c,j} = (d+\frac{j\tau}{m}, d+\frac{(j+1)\tau}{m}] \subset (d, D]$ with $j=0,...,m-1$, 
    \begin{equation}\label{Multi_Period_sol_c1}
        P^c(t)=P\left(\left(d+\frac{j\tau}{m}\right)^{+}\right)e^{-\mu_P \left(t-d-\frac{j\tau}{m}\right)}.
    \end{equation}
    For the sake of simplicity, we use the notation $\left(d+\frac{j\tau}{m}\right)^{+}=d+\frac{j\tau}{m}^{+}$. It follows from Eq.\,\eqref{Multi_Period_sol_c1} and the third equation of~\eqref{eq_many_predator} that
    \begin{equation*}
        P\left(d+\frac{(j+1)\tau}{m}^{+}\right)=P\left(d+\frac{j\tau}{m}^{+}\right)e^{-\mu_P \frac{\tau}{m}}+ \frac{\Lambda_P}{m}
    \end{equation*}
    which implies, by a recurrence argument on $j$, that
    \begin{equation*}
        P\left(d+\frac{j\tau}{m}^{+}\right)=P\left(d^{+}\right)e^{-\mu_P \frac{j\tau}{m}}+ \frac{\Lambda_P}{m}\sum_{i=0}^{j-1}e^{-\mu_P \frac{i\tau}{m}}, \forall j=1,...,m-1.
    \end{equation*}
    Using equation~\eqref{Multi_Period_sol_s} and the second equation of \eqref{eq_many_predator}, one has $P(d^+)=P(0^+)e^{-\mu_P d}+\frac{\Lambda_P}{m}$ that we replace in the above equation and obtain 
    \begin{equation*}
    \begin{aligned}
        P\left(d+\frac{j\tau}{m}^{+}\right)&= P\left(0^{+}\right)e^{-\mu_P \frac{j\tau}{m}}+\frac{\Lambda_P}{m}e^{-\mu_P \frac{j\tau}{m}}+ \frac{\Lambda_P}{m}\sum_{i=0}^{j-1}e^{-\mu_P \frac{i\tau}{m}},\\
        &=P\left(0^{+}\right)e^{-\mu_P \frac{j\tau}{m}}+ \frac{\Lambda_P}{m}\sum_{i=0}^{j}e^{-\mu_P \frac{i\tau}{m}},\\
        &=P\left(0^{+}\right)e^{-\mu_P\left(d+ \frac{j\tau}{m}\right)}+ \frac{\Lambda_P}{m}\left(\frac{1-e^{-\mu_P \frac{(j+1)\tau}{m}}}{1-e^{-\mu_P \frac{\tau}{m}}}\right)
    \end{aligned}
    \end{equation*}
    that holds for any $j=0,...,m-1$. By considering in particular the case $j=m-1$, 
    %
    we deduce from the fourth equation in~\eqref{eq_many_predator} that 
    \begin{equation}\label{exp_P_D}
    \begin{aligned} 
        P(D)&=e^{-\mu_P \frac{\tau}{m}}P\left(d+\frac{(m-1)\tau}{m}^{+}\right),\\
        &=P\left(0^{+}\right)e^{-\mu_P \tau}+ \frac{\Lambda_P e^{-\mu_P\frac{\tau}{m}}}{m}\left(\frac{1-e^{-\mu_P \tau}}{1-e^{-\mu_P \frac{\tau}{m}}}\right) .
    \end{aligned}
    \end{equation}
    
    Further, solving equation $\dot{P} = -\mu_P P$ for $ t\in (D,T]$ allows us to obtain the solution
    \begin{equation}\label{Multi_Period_sol_o} 
        P^o(t)=P(D^+)e^{-\mu_P(t-D)}
    \end{equation}
    which implies with the 
    fifth equation of~\eqref{eq_many_predator} that 
    \begin{equation*}
        P(T^+)=\phi_0P(D^+)e^{-\mu_P(T-D)}.
    \end{equation*}
    Since $P(D^+)=\phi_hP(D)$, we obtain thanks to Eq.\,\eqref{exp_P_D} 
    \begin{equation*}
    P\left(D^{+}\right)=\phi_h \left[P\left(0^{+}\right)e^{-\mu_P \tau}+ \frac{\Lambda_P e^{-\mu_P\frac{\tau}{m}}}{m}\left(\frac{1-e^{-\mu_P \tau}}{1-e^{-\mu_P \frac{\tau}{m}}}\right)\right]
    \end{equation*}
    from which we deduce that 
    \begin{equation*}
    \begin{aligned}
        P(T^+)=&\phi_0\phi_h  \left[P\left(0^{+}\right)e^{-\mu_P \tau}+ \frac{\Lambda_P e^{-\mu_P\frac{\tau}{m}}}{m}\left(\frac{1-e^{-\mu_P \tau}}{1-e^{-\mu_P \frac{\tau}{m}}}\right)\right]e^{-\mu_P(T-D)},\\
        =&\phi_0\phi_h  \left[ P\left(0^{+}\right)e^{-\mu_P T}+ \frac{\Lambda_P  e^{-\mu_P\frac{\tau}{m}}}{m}\left(\frac{e^{-\mu_P\left(T-D\right)}-e^{-\mu_P (T-d)}}{1-e^{-\mu_P \frac{\tau}{m}}}\right)\right].
     \end{aligned}   
    \end{equation*}
    The value for which $P\left(T^{+}\right)=P\left(0^{+}\right)$ is
    \begin{equation*}
        P\left(0^{+}\right)=  \frac{ \phi_0 \phi_h\Lambda_P}{m}\left[ \frac{e^{-\mu_P(T-D)}-e^{-\mu_P (T-d)}}{\left(1-\phi_0\phi_h e^{-\mu_P T}\right)\left(e^{\mu_P \frac{\tau}{m}}-1\right)}\right],
    \end{equation*}
    and then we can conclude that for $t\in (0, T]$, the m-cPDFS is given by $Y^T(t)=(0,0,P^T(t))$, with 
    \begin{equation}\label{Multi_Period_sol2_appendix}
        P^T(t)=\left\{
        \begin{aligned}
            &P^s(t)=P(0^+)e^{-\mu_P t},\; t\in I_s\\
            &P^c_{m,j}(t)=\left[P\left(0^{+}\right)e^{-\mu_P(d+ \frac{j\tau}{m})}+ \frac{\Lambda_P}{m}\left(\frac{1-e^{-\mu_P \frac{(j+1)\tau}{m}}}{1-e^{-\mu_P \frac{\tau}{m}}}\right)\right] e^{-\mu_P \left(t-d-\frac{j\tau}{m}\right)}, \; t\in I_{c,j} \\
            &P^o(t)=\phi_h\left[[P\left(0^{+}\right)e^{-\mu_P \tau}+ \frac{\Lambda_P e^{-\mu_P\frac{\tau}{m}}}{m}\left(\frac{1-e^{-\mu_P \tau}}{1-e^{-\mu_P \frac{\tau}{m}}}\right)\right]e^{-\mu_P (t-D)}; t \in I_o.
        \end{aligned}
        \right.
    \end{equation}
    with $I_s=(0,d]$, $I_{c,j}=(d+\frac{j\tau}{m}, d+\frac{(j+1)\tau}{m}]$, $j=0,\ldots,m-1$, $I_o=(D, T]$, and $P(0^+)$ being defined above.

\newpage
\bibliographystyle{unsrturl}
\bibliography{references}

\end{document}